# I came, I saw, I certified: some perspectives on the safety assurance of cyber-physical systems

Mithila Sivakumar, York University, Toronto, Ontario, M3J 1P3, Canada

Alvine B. Belle, York University, Toronto, Ontario, M3J 1P3, Canada

Kimya Khakzad Shahandashti, York University, Toronto, Ontario, M3J 1P3, Canada

Oluwafemi Odu, York University, Toronto, Ontario, M3J 1P3, Canada

Hadi Hemmati, York University, Toronto, Ontario, M3J 1P3, Canada

Segla Kpodjedo, Ecole de Technologie Supérieure, Montréal, Québec, H3C 1K3, Canada

Song Wang, York University, Toronto, Ontario, M3J 1P3, Canada

Opeyemi O. Adesina, University of Fraser Valley, Abbotsford, British Columbia, V2S 7M8, Canada

***Abstract*—** *The execution failure of cyber-physical systems (e.g., autonomous driving systems, unmanned aerial systems, and robotic systems) could result in the loss of life, severe injuries, large-scale environmental damage, property destruction, and major economic loss. Hence, such systems usually require a strong justification that they will effectively support critical requirements (e.g., safety, security, and reliability) for which they were designed. Thus, it is often mandatory to develop compelling assurance cases to support that justification and allow regulatory bodies to certify such systems. In such contexts, detecting assurance deficits, relying on patterns to improve the structure of assurance cases, improving existing assurance case notations, and (semi-)automating the generation of assurance cases are key to develop compelling assurance cases and foster consumer acceptance. We therefore explore challenges related to such assurance enablers and outline some potential directions that could be explored to tackle them.*

Cyber physical systems (CPS) are complex interoperable systems composed of interconnected and heterogenous components, with each component usually having its own set of inherent safety arguments[1, 2]. We now live in an era where the use of CPSs is constantly increasing, as illustrated by Amazon's plans to rely on revolutionary delivery drones to ensure the delivery of products of up to ten pounds within 30 mins. Artificial intelligence (AI)-enabled autonomous vehicles have gained unprecedented popularity across the world. Accordingly, several automobile manufacturers (e.g., Toyota, Tesla, Ford, Waymo, and General Motors) are now producing their own autonomous cars and realizing huge progress in the autonomy landscape[3]. Still, the execution failure of such safety-critical CPSs could potentially result in the loss of life, severe injuries, large-scale environmental damage, and major economic loss[4].

Ensuring safety of autonomous systems can therefore save lives, prevent injuries, reduce traffic, alleviate the costs associated with car accidents, and reduce the environmental damages caused by vehicles. Developing industry-wide safety standards and making sure the manufacturers of autonomous technologies comply with them is crucial to foster consumer acceptance. Producers of such CPSs are increasingly required to rely on assurance cases to demonstrate to regulatory authorities how they have complied with existing standards (e.g., ISO 26262)[5]. According to the SACM (Structured Assurance Case Metamodel) specification (https://www.omg.org/spec/SACM/2.2/About-SACM), an assurance case is a "*set of auditable claims, arguments, and evidence created to support the claim that a defined system/service will satisfy its particular requirements*". Assurance cases are a well-established structured technique used to document a reasoned, auditable argument supporting that a system meets some desirable properties, such as for safety or security. An assurance case comprises related collections of elements such as claims, arguments, and evidence. These elements are employed to establish that a given system will meet the specific properties of interest, such as safety or security requirements. Assurance cases are becoming very popular because they provide structured argumentation allowing to efficiently convey safety-critical information. Assurance cases are mainly used in safety-critical domains (e.g., automotive, healthcare, railways, aerospace) to deal with high-risk concerns and show to stakeholders that such systems are sufficiently safe according to domain-specific criteria. Figure 1 is adapted from '[Interlocking safety cases for unmanned autonomous systems in shared airspaces],'[6]. That Figure provides an excerpt of a safety assurance case represented in the GSN (Goal Structuring Notation) and developed for UAV (Unmanned Aerial Vehicle) Collision Avoidance. Details on GSN are available here: https://scsc.uk/gsn .

However, CPSs are complex due to the large number of their heterogenous components (e.g., mechatronic systems, sensors, actuators, software and networks) and relationships. This makes their safety assurance challenging, expensive, time-consuming, as well as labor-intensive[4, 5, 7]. Thus, to develop compelling assurance cases and foster consumer acceptance, several system assurance enablers should be explored. These include: the detection and mitigation of assurance deficits, the reliance on patterns to improve the structure of assurance cases and foster their reuse, the improvement of assurance case notations, and the (semi-)automation of the generation of assurance cases. It is therefore crucial to investigate challenges related to such assurance enablers and outline directions to address them that researchers and practitioners could explore. This could yield more advanced and efficient system assurance solutions focusing on safer CPSs. We focus on safety since it is a life-critical, non-functional requirement.

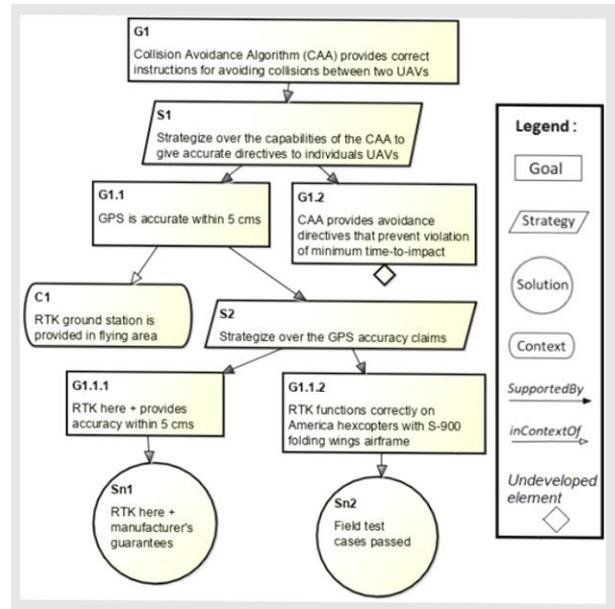

FIGURE 1. Partial safety case for UAV Collision Avoidance[6].

## LACK OF REUSE

Over the years, safety cases have been widely used to certify systems across various domains. This adoption has been supported in part by the introduction of safety case patterns.

Safety case patterns are generalized extensions of successful arguments — usually in the form of templates with placeholders for system-specific information — that can be re-used in the same or similar contexts to represent arguments that a system meets a safety property[1].

The application of technology to support and solve human needs across various domains requires independent systems to be interoperable, communicate seamlessly with other systems and thus leads to changes in their operating context[1]. "*In safety-critical domains, given a change in the system's operational context, the system's safety shall be re-assured.*"[1]. Hence, the manual creation or repeated manual modifications of safety cases for reassurance, and the domain knowledge required in the extraction and organization of system-specific artefacts due to a system's changing operating context, has posed challenges in the safety assurance of CPSs.

Numerous studies in the literature on safety case patterns[1, 2] emphasize the use of safety case patterns as a tool for reusability and reassurance. This supports two main objectives: (1) to keep track of all evidence and arguments generated under a system's operating context for traceability, which is checked against the current operating context during collaboration with another system[1]; (2) for automatic assurance case development based on automated collection of concrete information used as evidence and arguments contained in interconnected components that make up a CPS[2].

Still, the use of safety case patterns to promote reuse is challenging. Simple case studies are sometimes used to validate proposed safety case patterns (e.g., the partial construction of a safety case for a simplified airbag controller[1]), with a strong dependency on specific system design (e.g., model-based design[2]) and limited comparison in terms of assurance confidence level and associated cost when several patterns are appropriate for instantiation of a given claim. Hence, assurance case patterns should be applied to more complex case studies with a focus on safety-critical domains. Support for automatic system-specific evidence extraction for all CPSs, irrespective of their design method, is needed. Lastly, like coverage-based testing, where the goal is to utilize the fewest possible test cases to find the most defects, a tool or optimization approach in terms of cost effectiveness and assurance confidence level should be adopted to choose the best pattern when several patterns are applicable for instantiation of a given claim.

## RELATIVE LACK OF AUTOMATION

Safety case generation is typically manual, making the process expensive, error prone and labor-intensive[7]. It remains a challenge to understand, develop, assess, and maintain safety cases due to the large volume and diversity of information that a safety case should include when demonstrating compliance to relevant regulatory standards (e.g., ISO 26262, DO-178C). For example, the size of a preliminary safety case for surveillance on airport surfaces spans 200 pages[8]. And this is just preliminary and expected to grow further.

The need to automate or semi-automate the safety assurance process is emphasized in several studies[5, 9, 10]. For instance, Ramakrishna et al.[5] highlight how the increasing complexity of CPSs has made safety assurance case development more complicated. They stress that, although many research activities have been carried out in automating safety case instantiation and assembly, there is less research on automating the selection of safety case patterns. That automation could help expedite the creation of well-structured safety cases and foster their reuse. The ISO 26262 standard stipulates that the impact of system changes should be reflected in the safety case. But manually assessing and updating a safety case can introduce additional overhead, especially in agile development environments[9]. The manual creation of safety cases also presents challenges for internal safety certification and assessment processes[12]. Overall, the existing literature reinforces the need for automation or (semi-) automation in safety case development, management, and certification to mitigate the challenges posed by the growing complexity of CPSs.

A significant limitation identified in the existing literature is the lack of tool support for implementing safety case automation approaches[10]. Developing an entirely new tool can be a time-consuming task that requires substantial human resources. To overcome the limitations of tool support for safety case automation, researchers can adopt an approach that leverages existing tools like AdvoCATE or ExplicitCase, designed specifically for (semi-) automating the safety assurance process. These tools offer a variety of functionalities that can be extended and customized to accommodate new requirements. However, it is essential for the researchers to invest some time in exploring and studying the tools before attempting to modify or extend them. The goal is for them to gain sufficient knowledge about these tools and then adapt them to meet their own needs (e.g., automation of the creation and assembly of assurance arguments, and of the verification of safety cases). Moreover, when these tools are open source (e.g., ExplicitCase), researchers can delve into the code base and make necessary modifications to align them with their research objectives.

Noteworthy, to better support the automatic generation of safety cases, another interesting possible future avenue could be to explore the use of NLP (Natural Language Processing) and ML to extract (partial) assurance cases in a notation from free text.

## ASSURANCE DEFICITS

Assurance cases play a critical role in providing evidence of the safety and reliability of safety-critical systems. However, assurance deficits can occur when there are gaps (doubts) in the claims, inference rules, as well as in the evidence provided to support the claims made in the assurance case. Such deficits can contribute to uncertainty in general or to the occurrence of logical fallacies, as inadequate evidence or gaps in reasoning may lead to flawed arguments.

The challenges related to assurance deficits in safety-critical systems highlighted in the literature emphasize the difficulties in identifying, assessing, representing (e.g., using suitable modeling concepts), and mitigating uncertainties in assurance cases.

To address challenges related to assurance deficits — at least from a modeling perspective — the literature has explored some possible solutions. For instance, some approaches provide a comprehensive classification scheme for uncertainty monitoring techniques and understanding their relationship with the system architecture. This has the potential to enhance the ability to identify functional insufficiencies within the system architecture. Others have proposed solutions supporting the representation of uncertainties in safety arguments, thus extending the logical formalism of the GSN to consider all GSN components and refine the elicitation model to encourage more balanced and realistic uncertainty values. This has the potential to improve the accuracy and reliability of uncertainty representation in safety cases. Interestingly, some of the existing approaches have been augmented into a comprehensive framework that supports means to identify, prevent, tolerate (to some extent), remove, and predict uncertainty. Such frameworks can provide guidance for selecting suitable means to handle different types of uncertainty (e.g., aleatory, epistemic, and ontological types), thereby enhancing the overall dependability and safety of highly automated systems. By exploring such solutions, it is possible to address the challenges related to assurance deficits and improve the effectiveness of safety assurance processes in CPSs.

Interestingly, the literature has proposed a wide range of concepts (e.g., doubts, defeaters, uncertainties) to refer to assurance deficits and logical fallacies. However, there is no systematic effort that has been undertaken to make sure these expressions are properly categorized and used uniformly and consistently across research papers. Hence, existing representations of safety cases may fail to completely capture all the possible categories of assurance deficits and logical fallacies. As a result, existing techniques proposed to deal with (e.g., elicit/identify, assess, tolerate, mitigate) assurance deficits and logical fallacies may not cover all the possible assurance deficits and logical fallacies categories, and may therefore not be sufficiently accurate. This calls for the proposal of a taxonomy that categorizes the various types of assurance deficits and logical fallacies. Existing notations (e.g., GSN) will therefore build on that taxonomy to extend their concepts and better represent the different types of assurance deficits and logical fallacies as well as the associated relationships. This will facilitate the proposal of more advanced and efficient techniques to deal with the so-classified assurance deficits and logical fallacies.

## LACK OF FLEXIBILITY IN ASSURANCE CASE NOTATIONS

The inherent complexity of CPSs often makes structured arguments large and complicated. Hence, the arguments and evidence must be comprehensively documented to facilitate clear and defensible communication among diverse system stakeholders, including developers, reviewers, and regulators[11]. "*A convincing argument that is supported by evidence is the core of any assurance cases; therefore, they need to be clearly documented and represented*"[11].

There are three main ways to represent assurance cases. Free text, semi-structured text and more formal graphical representations that rely on one of the existing metamodels/languages/graphical notations such as SACM, GSN (Goal Structuring Notation), or CAE (Claim-Argument-Evidence). However, the use of fully formal notations is unmanageable in practice (e.g., in the industrial settings). So, although the use of GSN to represent assurance cases is very popular among researchers, a typical approach adopted in the industry is to represent the assurance case using free text. The use of textual representations is the easiest approach as it allows more flexibility to the developers and spares them from having the burden to learn the graphical representations (e.g., metamodels) mentioned above.

However, some problems arise when only text is used for representation. Free text may sometimes be ambiguous and difficult to understand as it may be poorly structured because not all engineers can write clearly and concisely. Also, the presence of multiple cross-references in free text may disrupt the flow of information. Unstructured textual representations may therefore lead to unreliable assurance case models that may be difficult to semantically verify[9]. Although state-of-the-art safety cases often involve claims expressed in free text form, the absence of formalized knowledge in machine readable format hinders automatic assessment of assurance case's consistency and semantics integrity[9]. Besides, in typically large assurance cases, it is also very challenging to navigate or browse through free text to find relevant information. It may therefore be difficult to show to the various system's stakeholders how exactly the argument structure assures safety requirements.

The challenges highlighted in literature[9,11] emphasize the need for the development of alternative assurance case representations that strike a balance between flexibility and verifiability. These new representations should provide the necessary flexibility to create complex arguments and, at the same time, enable semantics integrity and verifiability to ensure that the resulting assurance case can be trusted. By adopting such advancements, we can enhance the reliability and efficiency of the assurance process in a variety of application domains. Still, unless these new representations are fully formal, their use will necessarily heavily rely on heuristics and to some degree, textual analysis (e.g., NLP).

## SAFETY ASSURANCE OF ML-ENABLED SYSTEMS

Machine learning-based systems (especially those based on deep learning models) are gaining momentum in CPSs. Some popular applications of deep learning include autonomous systems such as autonomous driving systems and unmanned aerial systems.

Assuring the safety of ML-enabled autonomous systems is challenging for several reasons. To begin with, the uncertainty of the environment around the autonomous systems poses severe risk, as this may introduce unforeseen situations. Thus, ensuring the safety-critical decisions made by these systems are reliable and adhere to safety standards is challenging. A comprehensive approach needs to be adopted for addressing these challenges. For example, one may implement rigorous risk assessment, hazard and uncertainty analysis in the operating environment of autonomous vehicles. This requires carrying out an in-depth analysis of the system's behavior for both normal and exceptional cases. Another crucial aspect that can be investigated involves establishing robust verification and validation techniques to ensure the reliability of safety-critical decisions. This involves a thorough evaluation of the assurance case, implementing stringent testing protocols and assessing the system's performance in diverse conditions. By incorporating such approaches, the robustness of ML-enabled autonomous systems can be strengthened to effectively navigate uncertainties while ensuring reliable safety-critical decisions.

Although a variety of approaches address the assurance of ML-enabled systems, these techniques suffer from several limitations. For example, authors do not always empirically evaluate their approach by presenting case studies demonstrating that deep neural networks actually satisfy the claims. That limitation is a typical case in most studies in this domain. As a result, this issue limits the transparency of the assurance process in general. Thus, many stakeholders in the field are pushing to establish a standard procedure to deploy ML-based systems in scale production environments. Another key limitation of assuring ML-enabled systems — that is also affecting regular software — is the lack of formal verification techniques that can mathematically prove the correctness and safety of the ML models. Indeed, the growing inherent complexity and non-deterministic nature of ML models makes their formal verification challenging.

When proposing novel approaches focusing on ML-based safety case development, pattern generation, assessment and related topics, authors often refrain from presenting the complete safety case or providing a reference containing an entire safety case. This poses significant challenges to future researchers, especially to those who are interested in the safety assurance of ML-enabled systems. Without access to a complete safety case, deriving new methodologies for ML-based safety cases is exceedingly difficult. Furthermore, in contrast to traditional systems that achieve safety by relying on how developers specified safety requirements, ML-enabled systems can achieve safety by relying instead on specific examples chosen in a training data set. Still, since ML-enabled systems are not entirely explainable or comprehensible, it is challenging to assure that they took a safe decision for the right reasons. Such reasoning about safety is usually echoed in the literature when it comes to fairness in AI systems. Developing robust safety assurance solutions for ML-enabled systems therefore calls for the development of more explainable ML-enabled systems.

## REGULATORY COMPLIANCE OF ML-ENABLED SYSTEMS

Many CPSs are powered by ML. As reported in a recent article of the New York Times[12], Sam Altman (the CEO of Open AI) testified in a recent US hearing in which he explained the dangers and advantages of AI. He also urged lawmakers to regulate the AI systems produced in his company, as well as those of Google, Microsoft, and other tech companies. During his testimony, Mr. Altman stressed that collaborating with the government could help prevent the potential harm such systems could cause if they go wrong. Thus, in accordance with a suggestion Dr. Gary Marcus —a prominent emeritus professor from New York University and a critic of AI technology — made when testifying during the same hearing, Mr. Altman proposed the creation of a licensing agency that regulates AI systems. The agency's objective would be to issue licenses regulating the development of large AI models, safety regulations and tests that such models will have to pass prior to their release to end-users. That proposal emphasizing the need to regulate AI systems for purposes such as safety was well-received by the members of the Senate subcommittee. We therefore believe that moving forward the assurance of safety based on some standards is inevitable.

One of the main challenges with AI safety regulations is that it is unclear how lawmakers should proceed to formulate and enforce such regulations[12]. For instance, is there a need for national regulations or rather global regulations? Besides, is the enactment of stringent AI laws (that rightfully address safety risks) preventing important innovations in AI? It may pose threats to technological progress, and thus make a national economy less competitive than that of other countries with less stringent AI regulations. The other main challenge is that tech giants have typically fought the adoption of similar US bills (e.g., privacy, speech, safety bills), thus leading to the failure of their adoption under the form of US regulations[12]. The reluctance of such tech companies to adopt AI regulations may be due to concerns about keeping up with the tremendous technical progress characterizing AI systems. Hence, such regulations may make these companies liable for risks that should be tolerated due to the nature of such systems, but that may not have been anticipated by AI regulations and may therefore seem legally unacceptable.

Standards such as UL 4600 that focus on the safety of autonomous vehicles provide a framework for identifying and mitigating hazards[13]. But the automotive industry is not constrained to enforce them[13] and cannot therefore be held accountable in case of system failure. Creating laws that regulate AI while reflecting the constant technical progress achieved by companies is critical to fostering compliance in practice, thus making such companies accountable if the AI-enabled products they produce infringe public safety. Compliance with such regulations can be demonstrated using safety cases[13]. The rationale is that, according to the SACM specification, assurance cases allow the exchange of information among various system stakeholders (e.g., suppliers, acquirers), and between the operator and regulator, where knowledge regarding system requirements (e.g., safety, security) should be convincingly conveyed. Still, the understandability of the notations used to represent safety cases may be an additional challenge in this regard, especially if such notations go far beyond free text representations (e.g., metamodels, languages).

## TROUGH-LIFE SAFETY ASSURANCE

Dynamic safety cases (DSCs) have recently been introduced to assure through-life safety in various safety-critical domains (e.g., aerospace) and provide trusted autonomy to autonomous systems throughout their lifecycle[14]. Unlike traditional assurance cases, DSCs have runtime monitors allowing them to continuously assess system's requirements past the deployment of a system. DSCs support the operational assessment of assurance and ease intervention and fault-recovery in case change in operational data undermines assurance at run-time[14]. This is crucial for the certification of CPSs since the uncertainty of environments in which they operate may lead to complex and unanticipated risky situations. Such situations cannot be addressed by traditional assurance cases, which are only suitable prior to a system's deployment, and may become incorrect, obsolete or even inadequate during the system operation. DSCs therefore address the limitations of traditional assurance cases that need to be manually assessed and updated thus introducing additional overhead. However, dynamic safety assurance poses significant challenges that must be addressed to ensure the efficacy and trustworthiness of CPSs. One of such challenges is providing ongoing confidence in the efficacy of the assurance measures. The rationale is that approaches used to assess confidence and uncertainty in assurance cases mostly rely on popular mathematical theories or models such as Dempster-Schafer Theory (DST) or Bayesian analysis. However, these approaches usually allow computing confidence at design-time but not at run-time[15]. The confidence assessments made by these techniques may not reflect real-world situations where lack of sufficient evidence at run-time might undermine confidence expected at design-time and yield erroneous outcomes (e.g., fatal accidents)[15]. This calls for the development of new probabilistic assessment measures that address CPSs as the stochastic dynamic systems they usually are and that rely on random variables describing the state space of the CPSs[14].

## ACKNOWLEDGMENTS

This work is supported by a grant provided by Dr. Alvine B. Belle. We would like to warmly thank Dr. Lionel Briand for his precious feedback on this article.

**Mithila Sivakumar** is a research assistant at the Department of Electrical Engineering and Computer Science at York University, Toronto, Canada. Her research interests primarily revolve around the field of safety assurance, specifically related to topics such as safety cases, autonomous vehicles, automation, and machine learning. She holds a master's degree in computer engineering from the University of Texas, Dallas, USA. Contact her at msivakum@yorku.ca.

**Alvine B. Belle** is an Assistant Professor at the Department of Electrical Engineering and Computer Science at York University, Toronto, Canada. Her main research interests include software engineering, system assurance, machine learning, and EDI (Equity, Diversity, and Inclusion). She holds a PhD in software engineering from the University of Quebec (Ecole de Technologie Supérieure), Montreal, Canada. She has completed a 2-year industrial postdoctoral in software engineering at the University of Ottawa, Ottawa, Canada. She has recently completed a graduate diploma in public administration and governance at McGill University, Montreal, Canada. Contact her at alvine.belle@lassonde.yorku.ca.

**Kimya Khakzad Shahandashti** is a research assistant at the Department of Electrical Engineering and Computer Science at York University, Toronto, Canada. Her research interests primarily revolve around the field of safety assurance, specifically related to topics such as safety cases, assurance deficits, autonomous vehicles and automation. She holds a bachelor's degree in computer engineering from Shahid Beheshti University, Tehran, Iran. She is currently a master's student at the Department of Electrical Engineering and Computer Science at York University, Toronto, Canada. Contact her at kimya@yorku.ca

**Oluwafemi Odu** is a research assistant at the Department of Electrical Engineering and Computer Science at York University, Toronto, Canada. His research interests include the safety and security assurance of cyber-physical systems, and machine learning. He is currently a master's student at the Department of Electrical Engineering and Computer Science at York University, Toronto, Canada. Contact him at olufemi2@yorku.ca

**Hadi Hemmati** is an associate professor in the Department of Electrical Engineering and Computer Science, York University, Toronto, ON M3J1P3, Canada. He is also an adjunct associate professor at Calgary, AB, Canada. His main research interests include automated software engineering (with a focus on software testing, debugging, and repair) and trustworthy artificial intelligence (with a focus on robustness and explainability). His research has a strong focus on empirically investigating software/machine learning engineering practices in large-scale industrial systems. He is a Senior Member of IEEE. Contact him at hemmati@yorku.ca.

**Segla Kpodjedo** is an associate professor at the University of Quebec (Ecole de Technologie Superieure), Montreal, Canada. His current research interests include machine learning, software evolution, software engineering, graph matching, combinatorial optimization, and meta-heuristics. He received a Ph.D. degree in software engineering from Ecole Polytechnique de Montreal, Canada. Contact him at segla.kpodjedo@etsmtl.ca.

**Song Wang** is an Assistant Professor at the Department of Electrical Engineering and Computer Science at York University, Toronto, Canada. His research interests include software engineering, machine learning/deep learning, software reliability, software bug detection, software analytics, and crowdsourced testing. He holds a PhD in computer engineering from the University of Waterloo. Contact him at wangsong@eecs.yorku.ca.

**Opeyemi O. Adesina,** is an assistant professor at the School of computing of the University of Fraser Valley, Abbotsford, Canada. His current research interests include formal methods, software engineering, grid computing, artificial intelligence, and combinatorial optimization. He received a Ph.D. degree in computer science from the University of Ottawa. He completed a postdoctoral at the University of Waterloo, Canada. Contact him at opeyemi.adesina@ufv.ca.